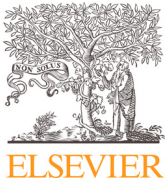
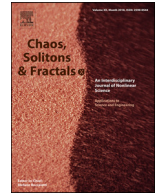

Frontiers

# New volatility evolution model after extreme events

Mei-Ling Cai [a], Zhang-HangJian Chen [a,b], Sai-Ping Li [c,d], Xiong Xiong [e,f], Wei Zhang [e,f], Ming-Yuan Yang [g,1], Fei Ren [a,h,*]

[a] *School of Business, East China University of Science and Technology, Shanghai 200237, China*
[b] *School of Economics, Anhui University, Hefei 230601, China*
[c] *Institute of Physics, Academia Sinica, Taipei 115, Taiwan*
[d] *Department of Physics, Hong Kong University of Science and Technology, Clear Water Bay, Kowloon, Hong Kong, China*
[e] *The College of Management and Economics, Tianjin University, Tianjin 300072, China*
[f] *China Center for Social Computing and Analytics, Tianjin University, Tianjin 300072, China*
[g] *China Academy of Financial Research, Zhejiang University of Finance and Economics, 8 Xueyuan Road, Hangzhou 310018, China*
[h] *Research Center for Econophysics, East China University of Science and Technology, Shanghai 200237, China*

A B S T R A C T

In this paper, we propose a new dynamical model to study the two-stage volatility evolution of stock market index after extreme events, and find that the volatility after extreme events follows a stretched exponential decay in the initial stage and becomes a power law decay at later times by using high frequency minute data. Empirical study of the evolutionary behaviors of volatility after endogenous and exogenous events further demonstrates the descriptive power of our new model. To further explore the underlying mechanisms of volatility evolution, we introduce the sequential arrival of information hypothesis (SAIH) and the mixture of distribution hypothesis (MDH) to test the two-stage assumption, and find that investors transform from the uninformed state to the informed state in the first stage and informed investors subsequently dominate in the second stage. The testing results offer a supporting explanation for the validity of our new model and the fitted values of relevant parameters.

© 2021 Elsevier Ltd. All rights reserved.

## 1. Introduction

In finance research, extreme events are normally referred to low probability and high impact events, such as occurrence of systemic risk and crisis in financial markets [1–3]. The study of volatility dynamics after extreme events can provide more accurate forecast for future volatility after large jumps, which is of utmost importance for risk management and portfolio optimization during crisis periods. Although the widely employed discrete time approaches like ARCH, GARCH models and continuous time models like stochastic volatility (SV) models have been devoted to describing the nature of volatility dynamics, a model that well describes the dynamic behavior of volatility after extreme events is still rare. This paper aims to fill this gap, and further the underlying mechanisms of volatility evolution will be discussed from the perspective of investor behaviors.

Empirical studies have shown that the statistical properties of volatility after extreme events are significantly different from the case of random fluctuations. Early works by [4] and [5] show that the volatility in the S&P500 after the endogenous event driven by large return shocks that come from within the system, e.g. the 1987 financial crash, reverts to the normal level more quickly than experience predicated and exhibit power law decay in a long time. [6,7] show that the volatility in UK, Japanese and US stock market after exogenous events induced by shocks of news that arrive from outside the system, such as the attack on September 11, 2001, also follows a power law decay. Further study later reveals that the volatility in the US stock market after exogenous and endogenous events both follow the power law decay, and the volatility after exogenous events decays faster than that after endogenous events [8]. The evolution of the volatility after experiencing a large intraday price change also follows a power-law decay for the stocks on the NYSE and NASDAQ [9].

Several types of dynamic models have been proposed to model the volatility after extreme events. [10] and [11] have proposed the random level shift (RLS) model, which can provide an accurate description of the level shifts in volatility of stock market after price jumps, but is not on the study of volatility itself. Other studies directly introduce the power law model to capture the decaying patterns of volatility after extreme events [5,8,9,12]. [6,7] introduce the Multifractal random walk (MRW) model [13], which unifies the properties of volatility multi-fractality and long-range dependence. The MRW model can provide a good fit to the volatility evolution after extreme events at time scales larger than 10 minutes or so. Reproducing the volatility evolution at small time scales is still a challenging task, which is of particular importance for portfolio selection and risk management in a short time immediately after extreme events.

* Corresponding author. Tel.: +86 021 64253369. School of Business, East China University of Science and Technology, 130 Meilong Road, Shanghai 200237, China
*E-mail address:* fren@ecust.edu.cn (F. Ren).
[1] Phone: +86 0571 86735252.

The stochastic volatility (SV) models constructed based on stochastic process, such as Heston model [14], Hull-White model [15], stein and stein model [16], 3/2 model [17], 4/2 model [18], and SV model with jumps [19,20] are mainly dedicated to the study of volatility dynamics in options markets. However, relatively little attention has been paid to the problem of volatility dynamics in equity markets by using this model. The SV models can well describe the volatility distribution over a relatively wide volatility range, including lognormal model and Hull-White model [21], exponential Ornstein-Uhlenbeck (expOU) model [22], and Heston model and Hull-White model [23]. Further extensions incorporate stylized facts like volatility autocorrelation and leverage effect [22], multiple time scales [24,25], and price stability [26]. This type of model may not be directly used to model the volatility dynamics after extreme events [8], but it offers a good candidate for its success in describing stochastic feature of volatility fluctuations in stock markets with wide applicability.

To describe the dynamical behavior of volatility after extreme events over the entire time domain, we here propose a new dynamical model based on the SV model. Since there exist large volatility jumps after extreme events, we therefore introduce the Dirac delta function to represent the instantaneous impact of these events [27,28]. The mean-reverting factor in SV model can well describe the tendency of volatility to revert to the long-run equilibrium value. Accordingly, we retain the mean-reverting factor, and use it to model the reverting process to equilibrium after large jumps. Inspired by the work [29], in which a two-region model of stochastic volatility with different forms of reverting factor are proposed to describe the lognormal pdf for low volatility values and a power-law pdf for large volatility values, we introduce a two-stage volatility evolution model with different reverting factors to capture the volatility dynamics after extreme events for both small and long time scales. Our work is also in line with the works of [24,25,30,31], which provide valid supports for the consideration of different diffusion processes for short and long time scales.

Differing from the SV model, the mean-reverting factor in our model is inversely proportional to time by referring to the time-dependent friction in the Langevin equation [32], which has been widely used to describe the dynamics of stock price fluctuations [33–36]. Due to the wide divergence in investors' ability of acquiring information and the great changes in their trading behaviors, we infer that the mean reversion rate in the mean-reverting factor may be time-dependent. By making an analogy between the friction that slows down the speed of particles in fluid and drives the system back to equilibrium, and the mean-reverting factor that has a mean reversion rate decreasing over time and drives the volatility reverting back to the long-run equilibrium value, we follow the work of [32] and assume the mean reversion rate is inversely proportional to time in a power-law form. The power exponent change for small and large time scales in different stages is in parallel with the ideas in [24,25,29–31].

To further demonstrate the descriptive power of our new model, we also use this model to describe the empirical results of volatility evolution after endogenous and exogenous events. By fitting the volatility evolution after endogenous events with our model, we find that the volatility shows a stretched exponential decay in the first stage $t \leq 10$ minutes and a power law decay in the second stage $t > 10$ minutes. While for the exogenous events, there only exists the second stage when the volatility follows a power law decay. The results show that our model provides a very good fit to the volatility evolution after both endogenous and exogenous events. As mentioned above, the MRW can well capture power law decay of volatility for $t > 10$ minutes after both endogenous and exogenous events [6,7], and the analytical result is difficult to obtain. In comparison, our model can well describe the volatility behavior within 10 minutes after endogenous events with stretched exponential decay, when the accurate prediction of volatility in the initial stage after extreme events is of particular importance for portfolio selection and risk management. The rapid change of volatility in this period contains rich information, and investors' receiving of these information is generally completed within 10 minutes after extreme events, which will be verified in the assumption test in Section 4.3. These make the applications of volatility prediction in initial time more efficient through the timely adjustment of investors' trading strategy and policies implemented by regulators.

Different from the previous SV models focused on the mathematical equations and their analytical solving procedures, we perform a two-stage assumption test to explore the underlying mechanisms under different stages. This further provides a rational support for our two-stage model from empirical studies. We introduce the sequential arrival of information hypothesis (SAIH) [37,38] and the mixture of distribution hypothesis (MDH) [39–41], and propose a two-stage assumption that investors sequentially transform from the uninformed state to the informed state in the first stage and informed investors subsequently dominate the market in the second stage. The testing results show that volatility evolution after extreme events can be separated into two stages by a time constant $t_w$, which represents the time needed for investors to complete the transformation from uninformed to informed. The power exponent in the mean-reverting factor for the stages $t \leq t_w$ and $t > t_w$ varies due to the fact that investors have different behaviors in different stages. This can help us understand the two stage scenario and functional form of the response to impact in our model. We believe that our study provides a unified framework for understanding and explaining the underlying mechanism of volatility dynamics after extreme events from the microscopic level of investor behaviors.

Compared to previous studies, our paper makes several contributions summarized as follows. First, we propose a new dynamic model which can accurately describe the two-stage decay patterns of volatility after extreme events over the entire time domain. Second, our new model can successfully describe and distinguish the empirical results of volatility evolution after endogenous and exogenous events, which makes our new model flexible and suitable for practical applications. Third, we explore the underlying mechanism of volatility evolution after extreme events. By testing the hypotheses of SAIH and MDH, we find the investors transform from uninformed state to informed state in the first stage and informed investors dominate the market in the second stage.

The rest of this paper is organized as follows. Section 2 presents the methodology. Section 3 describes the data. Section 4 presents and discusses the empirical results, and gives the robustness and universality tests. We conclude in Section 5.

## 2. Methodology

In this section, we present a method to identify extreme events, including endogenous and exogenous events, in order to obtain empirical results of volatility dynamics after extreme events. We will propose a new dynamical model, the volatility evolution (VE) model, to describe volatility evolution after extreme events. We will then introduce two hypotheses, namely SAIH and MDH, and propose the two-stage assumption for exploring the mechanisms underlying volatility evolution after extreme events in different stages.

### 2.1. Identification of extreme events

It is well known that an extreme event with high impact, such as occurrence of systemic risk and crisis in financial markets, asset price and volatility respond quickly and substantially to these extreme event [42]. Therefore, we identify extreme events by seeking for large jumps in volatility, which satisfy the condition $V(t) > S\langle V(t)\rangle_d$, where $V(t)$ is the volatility at time $t$, $\langle V(t)\rangle_d$ is the mean value of $V(t)$ for all trading days, and the threshold $S$ is a positive constant [5,8]. The simplest and widely used measurement of price volatility is tracking the absolute return values, especially for the high frequency time series [22,43–45]. Hence, in this paper we introduce the volatility as the absolute value of the logarithmic price return $R(t)$, which is given by

$$R(t) = \ln(P(t+1)) - \ln(P(t)), \quad (1)$$

where $P(t)$ is the price at time $t$.

Next, we classify the extreme events into two types, namely endogenous and exogenous events. Following the studies of [46], [47] and [48], the exogenous events can be defined as the events synchronized with the news release of stocks. Specifically, we identify the exogenous events as events that happen within one minute center around news are released. The endogenous ones are the events driven by large return shocks that come from within the system, and are usually not accompanied by news that arrive from outside the system. Therefore, we simply define the endogenous events as the extreme events that exclude the exogenous events [8,46].

### 2.2. The VE model

The VE model proposed here is based on the SV model, in which the volatility $V(t)$ can be expressed as

$$dV(t) = b(\omega - V(t))dt + \eta(t), \quad (2)$$

where $b$ is the mean reversion rate, i.e., the rate that the volatility reverts back to a long-time level $\omega$, and $\eta(t) = \sigma_v dW$ represents the stochastic fluctuation of volatility, where $\sigma_v$ is the standard deviation of $V(t)$ and $dW$ is a standard Wiener process with zero mean.

To model the volatility evolution after the impacts of extreme events, one need to model the instantaneous impact of extreme events first. We introduce the Dirac delta function in the SV model to represent the instantaneous impact of extreme events $F_e(t) = \rho\delta(t)$ [27,28], where $\rho$ denotes the impact strength, and $\delta(t)$ is the Dirac delta function given by

$$\delta(t) = \begin{cases} 0 & t \neq 0 \\ +\infty & t = 0 \end{cases}. \quad (3)$$

The Dirac delta function, $\delta(t)$ is infinite at $t = 0$, the instant when the impact of extreme event occurs, and zero elsewhere. This feature makes it possible to represent the instantaneous impact of an extreme event at the initial time.

After impacts of extreme events, the volatility tends to revert to the long-time mean value. Considering the $b(\omega - V(t))$ in the SV model can well describe the tendency of volatility to revert to the long-run equilibrium value, where $b$ is the mean reversion rate, i.e., the speed of mean reversion, we therefore retain the mean-reverting factor in the VE model. Based on the definition of mean-reverting factor in the SV model [20], the change of volatility $dV(t)$ is proportional to $-V(t)$, hence the mean-reverting factor $F_r$ takes the form $F_r = -bV(t)$ when the long-run equilibrium value $\omega \approx 0$[1]. The volatility in our VE model is strictly greater than or equal to zero, which is consistent with the classical SV model [14]. Differing from the SV model, the mean-reverting factor in our model is inversely proportional to time by referring to the time-dependent friction in the Langevin equation [32], which has been used to describe the dynamics of stock price fluctuations [33–36]. Due to the wide divergence in investors' ability of acquiring information, the investors' trading behaviors can change over time, and we therefore infer that the mean reversion rate may be time-dependent. Taking into account the condition when $t \to \infty$, the mean reversion rate will go to zero. Hence the mean reversion rate decreases over time, similar to the restoring force in the form of $1/\lambda(t)$ in Ornstein-Uhlenbeck process [31]. The friction in the Langevin equation [49] slows down the speed of particles in fluid and drives the system back to the equilibrium state, which is analogous to the mean-reverting factor that has a reversion speed decreasing over time and drives the volatility to revert to the long-run equilibrium value. We thus use the power-law decaying form of the friction proposed in [32], and assume that the mean-reverting factor takes the form $F_r(t) = -\frac{b}{t^c}V(t)$ here.

As shown in the empirical results and the results of the assumption tests proved in Section 4 below, the volatility decay after extreme events displays two stages, similar to the two-region volatility PDF in [29]. We therefore introduce the time constant $t_w$ to separate the two stages, namely $t \leq t_w$ and $t > t_w$, where $t_w$ represents the time needed for investors to complete the transformation from uninformed to informed. The parameter $c$ reflects the changes of investors' information acquisition and decision-making behavior correspondingly over time $t$, and is assumed to have different values in two stages, which leads to different speeds that volatility reverts to equilibrium after impacts. This is in line with the works with different diffusion processes for short and long time scales [24,25,30,31]. The setting of parameter $c$ for different stages is as follows. In the first stage $t \leq t_w$, investors transform from uninformed state to informed state. Hence the informed investors and their decision-making behaviors increase over time, which directly leads to the increase of investors' trading. Since there exists the causality relation from trading volume to price changes [50,51], the market becomes more volatile, which in turn gives rise to the increase of the resistance to recovery, and the mean-reverting factor would decrease slowly as time $t$ increases. Considering that the mean-reverting factor $F_r \propto 1/t^c$, we here assume $c > 1$ to ensure the slow decrease of $F_r$ in this stage. In the second stage when $t > t_w$, the process in which investors transform from the uninformed state to the informed state is complete, and there is no increase of informed investors. This leads to the decrease of decision-making behaviors and their trading behaviors. Thus the market volatility tends to be restrained, which then causes the decrease of the resistance to recovery, and the mean-reverting factor would decrease faster as time $t$ increases. In a similar fashion, we assume that $c = 1$ to ensure the fast decrease of $F_r$ in this stage. Other parameters and factors in our model have the same values in different stages.

Finally, substituting the instantaneous impact $F_e(t) = \rho\delta(t)$ and the mean-reverting factor $F_r(t) = -\frac{b}{t^c}V(t)$ into the SV model (Eq. (2)) gives

$$dV(t) = \rho\delta(t)dt - \frac{b}{t^c}V(t)dt + \eta(t). \quad (4)$$

---

[2] If the volatility is defined as the standard deviation of stock return, the long-run equilibrium level would be greater than 0. Hence, we also give the solution and results of the VE model with a long-run equilibrium level $\omega > 0$ in Appendices A.2 and B.1, respectively. The results indicate that the VE model with $\omega = 0$ is optimal with respect to our definition of volatility.

This equation can describe the dynamical motion of volatility after an instantaneous impact on the stock market. Using linear response theory, Eq. (4) can be solved exactly (see appendix A for the detailed derivation) and the average solution is

$$\langle V(t) \rangle = \begin{cases} \rho e^{\frac{bt^{1-c}}{c-1}} & t \leq t_w \\ \rho t^{-b} & t > t_w \end{cases}. \qquad (5)$$

### 2.3. Hypotheses of SAIH and MDH

To explore the underlying mechanism that drives the volatility evolution in different stages, we propose a two-stage assumption that investors transform from the uninformed state to the informed state in the first stage and informed investors subsequently dominate the market in the second stage. The assumptions in these two stages are proposed based on the SAIH and MAD hypotheses respectively, and the detailed testing methods are given as follows.

#### 2.3.1. SAIH

The SAIH proposed by [37] and [38] argues that the formation of new market equilibrium is not instantaneous and requires some time for investors to react to new information and make trading decisions. Once all investors have reacted to the information signal and made their trading decisions, a final equilibrium is reached. This results in a lead-lag relationship between information flow and volatility [52].

The lead-lag relations between trading volume, which is considered as the proxy for information flow, and volatility are generally examined by using Granger-causality test to verify the SAIH [52–54]. Following these studies, we construct the Granger-causality model [55] as follows

$$V(t) = \gamma_1 + \sum_{k=1}^{p} g_k V(t-k) + \sum_{k=1}^{p} h_k Z(t-k) + o_1(t), \qquad (6)$$

$$Z(t) = \gamma_2 + \sum_{k=1}^{q} l_k V(t-k) + \sum_{k=1}^{q} m_k Z(t-k) + o_2(t), \qquad (7)$$

where $V(t)$ is the volatility, and $Z(t)$ is the logarithm of trading volume, $\gamma_1$ and $\gamma_2$ are the intercepts of Eqs. (6) and (7), $g_k$ and $m_k$ are the first-order autoregressive coefficients, $h_k$ and $l_k$ measure the lead-lag relations between trading volume and volatility, $o_1(t)$ and $o_2(t)$ are the regression errors, $p$ and $q$ are the optimal lag lengths obtained by using the Akaike's Information Criterion (AIC).

Linear casual relationships can be inferred from Eqs. (6) and (7). If $h_k = 0$ for all $k$, then Eq. (6) implies that past trading volume has no influence on the volatility, i.e., trading volume does not linearly Granger cause volatility. Similarly, volatility does not linearly Granger cause trading volume if $l_k = 0$ for all $k$ in Eq. (7). On the other hand, if $h_k \neq 0$ ($l_k \neq 0$) at the 5% significance level for some values of $k$, it will imply that trading volume (volatility) linearly Granger causes volatility (trading volume). As predicted by the SAIH, the sequential reaction of investors to information produces a lead-lag relationship between the volatility and trading volume during the equilibrium achieving procedure. Therefore, if there is a significant lead-lag relationship between the volatility and trading volume, there will exist the stage that investors sequentially transform from uninformed state to informed state after extreme events.

#### 2.3.2. MDH

The MDH was first proposed by [39] and modified by [40] and [41] to explain the volume-volatility relations of stocks. According to the MDH [56,57], the trading volume can generally be divided into informed and uninformed components. The specific steps of classifying the trading volume are as follows.

First, take the logarithm of the trading volume and obtain a new volume series $Z(t)$.

Second, remove the time trend of intraday volume by using the linear regression model with time trend variables $t$ and $t^2$, which is given by

$$Z(t) = a_0 + a_1 t + a_2 t^2 + \varphi(t), \qquad (8)$$

where the residual $\varphi(t)$ is the trading volume after removing the time trend. There are two trading sessions of Shanghai Stock Exchange, i.e., morning and afternoon sessions, and the intraday trading volume shows double $U$-shaped pattern in the Shanghai stock market [58]. For these reasons, we remove the time trend of trading volume by using Eq. (8) for the morning and afternoon trading sessions separately.

Third, divide the detrended trading volume $\varphi(t)$ into expected and unexpected volume using ARMA model (Auto-Regressive and Moving Average model), which is given by

$$\varphi(t) = c_0 + \sum_{i=1}^{p} \theta_i \varphi(t-i) + u(t) + \sum_{j=1}^{q} \phi_j u(t-j), \qquad (9)$$

where $p$ is the dimension of the autoregressive component AR($p$) given by $\varphi_{AR}(t) = c_0 + \sum_{i=1}^{p} \theta_i \varphi(t-i)$, and $q$ is the dimension of the moving average component MA($q$) given by $\varphi_{MA}(t) = u(t) + \sum_{j=1}^{q} \phi_j u(t-j)$, and $u(t)$ is the residual from AR($p$). Fitted value $\hat{\varphi}(t)$ from Eq. (9) represents the expected volume, and residuals $\varphi(t) - \hat{\varphi}(t)$ represents the unexpected volume. The uninformed volume $Z_u(t)$ is the predictable component of trading volume, and can be measured as $Z_u(t) = \hat{\varphi}(t)$. The informed volume $Z_i(t)$ induced by information arrival cannot be explained by historical series of trading volume, but can be measured as $Z_i(t) = \varphi(t) - \hat{\varphi}(t)$.

To identify which component of the trading volume has significant influence on the volatility, we will perform another linear regression analysis on the volatility-volume relation test, which is expressed as

$$V(t) = \beta_0 + \beta_i Z_i(t) + \beta_u Z_u(t) + \varepsilon(t) \qquad (10)$$

where $V(t)$ is the volatility, $Z_i(t)$ and $Z_u(t)$ denote the informed and uninformed trading volumes respectively. If the coefficient $\beta_i$ ($\beta_u$) is significant and $\beta_u$ ($\beta_i$) is insignificant, it will imply that volatility $V(t)$ is mainly influenced by the informed (uninformed) volume $Z_i(t)$ ($Z_u(t)$), and thus the informed (uninformed) investors will dominate the market.

### 3. Data

To study the volatility evolution of stock market after extreme events and explore the underlying mechanism of volatility evolution, we first introduce the dataset which records the closing prices and the trading volumes of the SSE 180 index, a benchmark index for the Shanghai market, at the end of every minute (from 9:30 a.m. to 11:30 a.m. and from 13:00 p.m. to 15:00 p.m.) for 717 consecutive trading days from December 16, 2013 to November 22, 2016. To ensure that our findings in this paper is robust, we also employ the data of closing prices of the SSE 180 index at the end of every minute for 119 consecutive trading days from December 12, 2016 to June 11, 2017 for the robustness test.

The next dataset records the news for all constituent stocks of the SSE 180 index reported on the website of SinaFinance (http://finance.sina.com.cn/), which is one of the top financial websites in China, also from December 16, 2013 to November 22, 2016. We identify a total of 173,419 relevant news and record the time in minutes that these news released. News for the SSE 180 index, which is a capitalization-weighted average of the prices of all its constituents, per se is rare. We therefore collect the news for all constituent stocks of the SSE 180 index, since the news related to the constituent stock may have impacts on the stock itself as well as the index. This would provide a more comprehensive collection of news for the SSE 180 index. We identify the exogenous events as extreme events that happen within one minute center around the release of news.

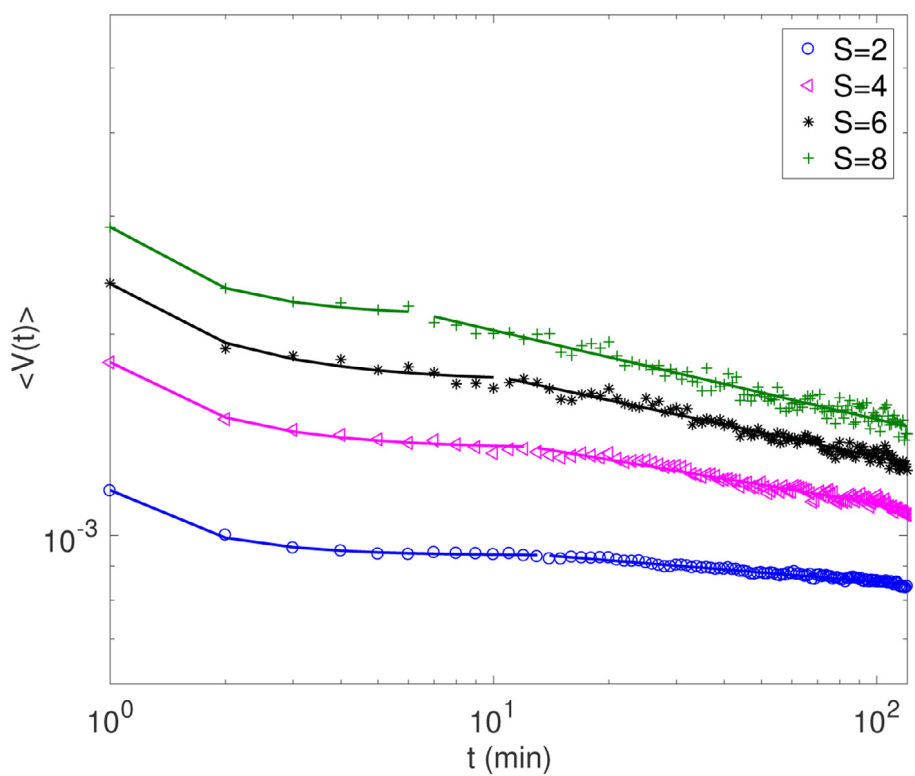

**Fig. 1.** Volatility evolution after extreme events for threshold $S = 2, 4, 6, 8$ using the data of the SSE 180 index from December 16, 2013 to November 22, 2016. The solid lines in the figure are best fits to the data using Eq. (5).

For the universality test, we introduce the dataset which records closing prices of HSI index in the Hong Kong Exchange at the end of every minute (from 09:30 a.m. to 12:00 a.m. and from 13:00 p.m. to 16:00 p.m.) for 132 consecutive trading days from January 11, 2016 to July 25, 2016.

## 4. Empirical results and discussion

In this section, we present the empirical results of volatility evolution of the SSE 180 index after extreme events, including exogenous and endogenous ones, for the period from December 16, 2013 to November 22, 2016. We then describe the empirical results of volatility evolution with the VE model proposed in Section 2.2. We finally explore the underlying mechanism of volatility evolution after extreme events by testing the two-stage assumption proposed based on the SAIH and MDH hypotheses, and give explanations for our new model from the perspective of investor behaviors.

### 4.1. Empirical results of volatility evolution after extreme events

By identifying extreme events with the method in Section 2.1, we here present the volatility evolution of the SSE 180 index after extreme events with different impact strength for threshold $S = 2, 4, 6, 8$ in Fig. 1. The volatility $\langle V(t) \rangle$ is taken average over the events with a specific value of threshold $S$. Notice that the larger threshold $S$ is, the greater the event impact will be. For different values of threshold $S$, the empirical results in show that the decays of volatility have similar patterns.

From, one can see that the decays of volatility are close to power laws at later times, which is consistent with previous studies [5,8,9], and one can also see that the volatility decays deviate from the power laws in the initial stage of the decays. These results indicate that the evolution behavior of volatility may differ in different stages after extreme events.

It is well known that events in stock markets can be classified into two different types, namely exogenous events and endogenous events. These two types of events have significantly different effects on the stock volatility [8,27]. To further study the influence of events on the stock market volatility, we present the empirical results of volatility evolution after endogenous and exogenous events respectively.

By identifying the extreme events with the method proposed in Section 2.1, we calculate the number of extreme events, including endogenous and exogenous events for the SSE 180 index, and present the results in Fig. 2. One can see that the number of all extreme events is slightly larger than the endogenous events and significantly larger than the exogenous events for the index. This suggests that most of the extreme events are endogenous.

Next, we present the volatility evolution of the SSE 180 index after endogenous and exogenous events for threshold $S = 2, 4, 6, 8$ in Fig. 3. One can observe that in the figure that the decays of volatility for different values of threshold $S$ have similar patterns. In Fig. 3(a), the volatility decay after endogenous events deviates from the power laws in the initial stage and is close to power law behavior at a later stage, which is consistent with the results in since most of the extreme events are endogenous. By contrast, the results in Fig. 3(b) show that the volatility after exogenous events obeys a power law from the beginning for different values of threshold $S$.

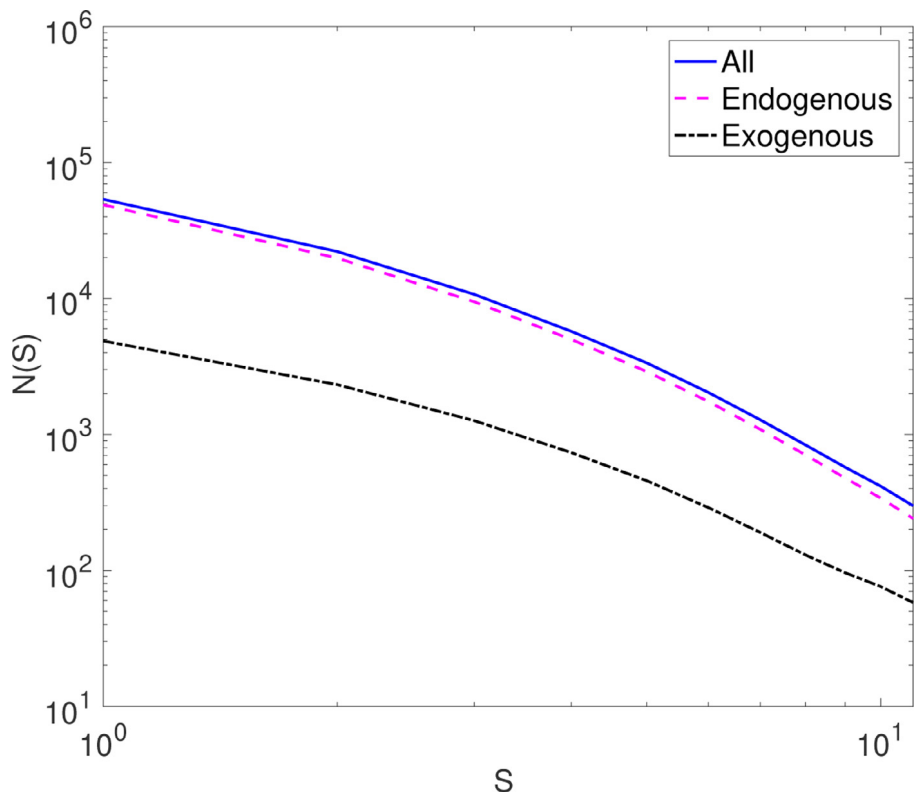

**Fig. 2.** Numbers of all extreme events, endogenous events and exogenous events as a function of threshold $S$ for the SSE 180 index during the period from December 16, 2013 to November 22, 2016.

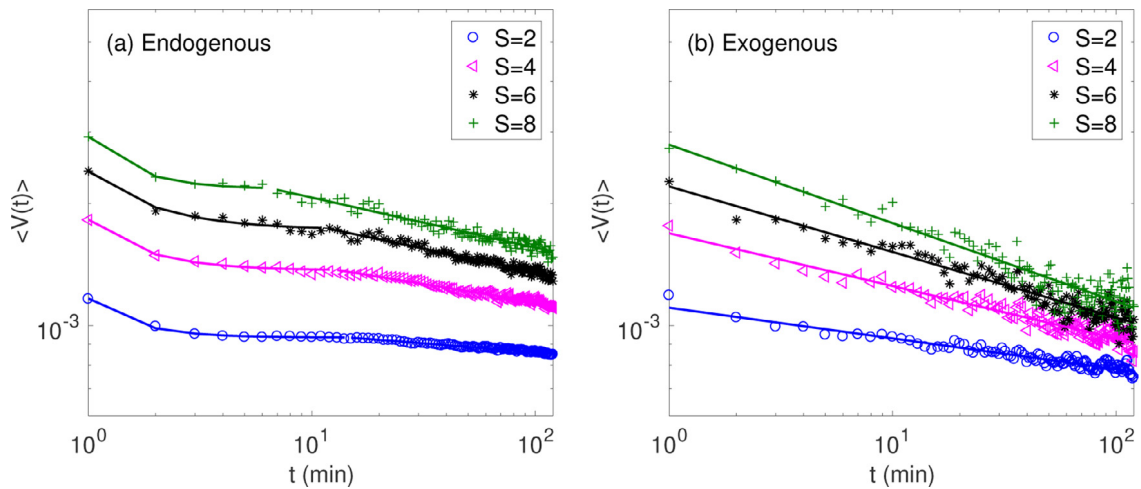

**Fig. 3.** Volatility evolution after (a) endogenous and (b) exogenous events for threshold $S = 2, 4, 6, 8$ using the data of the SSE 180 index from December 16, 2013 to November 22, 2016. The solid lines in the figure are best fits to the data with Eq. (5).

*4.2. Fitting volatility evolution after extreme events with VE model*

In the following, we will fit the volatility evolution after extreme events with Eq. (5) using the least-squares approach, and plot the best fitted curves in. We use the non-linear least square method to fit the curves based on the Levenberg-Maruardt algorithm [59]. We estimate the RSS under the different values of parameteres $\rho$, $b$ and $c$ with the initial value of zero and $t_w$ with the initial value of 2, and RSS is defined as the sum of squares of the difference between solutions and actual values of volatility in the two regions $t \leq t_w$ and $t > t_w$. The optimal parameters of $t_w$, $\rho$, $b$ and $c$ are determined by the minimization of RSS. The corresponding estimated parameters of the best-fits are given in Table 1.

In, one can observe that the volatility evolution after extreme events can be well fitted by a stretched exponential decay in the first stage and a power law decay in the second stage. As shown in Table 1, the RSS of all best-fits for different thresholds have extremely small values that are close to zero (the largest value is $5.67 \times 10^{-9}$). This means that our model provides well fits to the volatility evolution after events with different impact strength[2]. Furthermore, as the threshold $S$ increases, the impact strength $\rho$

---

[3] We also perform the parameter estimates of VE mode for different scenarios in which the volatility evolution obeys stretched exponential or power law in two different stages. The results show that the values of RSS are smallest when the volatility evolution after endogenous events follows a stretched exponential decay when
6

**Table 1**

Parameter estimates of VE model for the volatility evolution of the SSE 180 index after extreme events for the period from December 16, 2013 to November 22, 2016.

| S | $t_w$ | $\rho e^{\frac{bt^{1-c}}{c-1}} (t \leq t_w)$ | | | $\rho t^{-b} (t > t_w)$ | | RSS |
|---|---|---|---|---|---|---|---|
|   |   | $\rho$ | $b$ | $c$ | $\rho$ | $b$ |   |
| 2 | 13 | $0.89 \times 10^{-3}$ | 0.41 | 2.81 | $1.06 \times 10^{-3}$ | 0.05 | $3.17 \times 10^{-10}$ |
| 4 | 12 | $1.26 \times 10^{-3}$ | 0.42 | 2.37 | $1.73 \times 10^{-3}$ | 0.10 | $1.45 \times 10^{-9}$ |
| 6 | 10 | $1.65 \times 10^{-3}$ | 0.43 | 2.08 | $2.31 \times 10^{-3}$ | 0.12 | $2.49 \times 10^{-9}$ |
| 8 | 6  | $2.03 \times 10^{-3}$ | 0.46 | 1.88 | $2.75 \times 10^{-3}$ | 0.13 | $5.67 \times 10^{-9}$ |

This table reports the best-fit values of VE model as described by Eq. (5) for the volatility evolution of SSE 180 index after extreme events. The last column reports the values of the residual sum of squares (RSS) for the best-fit of our VE model.

**Table 2**

Parameter estimates of VE model for the volatility evolution of the SSE 180 index after endogenous and exogenous events for the period from December 16, 2013 to November 22, 2016.

|  | S | $t_w$ | $\rho e^{\frac{bt^{1-c}}{c-1}} (t \leq t_w)$ | | | $\rho t^{-b} (t > t_w)$ | | RSS |
|---|---|---|---|---|---|---|---|---|
|  |   |   | $\rho$ | $b$ | $c$ | $\rho$ | $b$ |   |
| Endogenous | 2 | 14 | $0.93 \times 10^{-3}$ | 0.45 | 3.00 | $1.05 \times 10^{-3}$ | 0.04 | $2.20 \times 10^{-9}$ |
|  | 4 | 12 | $1.36 \times 10^{-3}$ | 0.46 | 2.57 | $1.73 \times 10^{-3}$ | 0.09 | $2.49 \times 10^{-8}$ |
|  | 6 | 11 | $1.70 \times 10^{-3}$ | 0.43 | 2.26 | $2.32 \times 10^{-3}$ | 0.12 | $8.99 \times 10^{-8}$ |
|  | 8 | 6  | $2.14 \times 10^{-3}$ | 0.55 | 2.09 | $2.75 \times 10^{-3}$ | 0.13 | $2.28 \times 10^{-7}$ |
| Exogenous | 2 | 0 |  |  |  | $1.11 \times 10^{-3}$ | 0.08 | $4.29 \times 10^{-8}$ |
|  | 4 | 0 |  |  |  | $1.69 \times 10^{-3}$ | 0.13 | $1.72 \times 10^{-7}$ |
|  | 6 | 0 |  |  |  | $2.20 \times 10^{-3}$ | 0.16 | $4.48 \times 10^{-7}$ |
|  | 8 | 0 |  |  |  | $2.79 \times 10^{-3}$ | 0.19 | $1.03 \times 10^{-6}$ |

This table reports the best-fit values of VE model as described by Eq. (5) for the volatility evolution of SSE 180 index after endogenous and exogenous events. The last column reports values of the residual sum of squares (RSS) for the best-fit of our VE model.

and parameter $b$ increase, but the parameters $t_w$ and $c$ decrease. We also conduct a comparison test between the VE model and the Heston model [14]. Table B.9 in Appendix B.1 presents the results of parameter estimates of the Heston model, and the results show that the VE model better describe the evolution of volatility after extreme events.

It was shown in Section 4.1 that the dynamical behaviors of volatility after endogenous and exogenous events are significantly different, hence we will introduce the fitted results of our VE model for volatility evolution after the two types of events separately in the following. The volatility evolutions after endogenous and exogenous events are fitted with Eq. (5) using the least-squares approach, and the corresponding estimated parameters of the best-fits are given in Table 2. As shown in Table 2, the RSS of all best-fits for different thresholds have extremely small values. For instance, the largest value is $2.28 \times 10^{-7}$ and $1.03 \times 10^{-6}$ for endogenous and exogenous events respectively, suggesting that our model provides well fits to the volatility evolution after both types of events with different impact strength. More specifically, the volatility evolution after endogenous events follows stretched exponential decay when $t \leq t_w$ and power law decay when $t > t_w$, while the volatility evolution after exogenous events follows power law decay for the entire time domain. For the tendency of estimated parameters $t_w$, $\rho$, $b$ and $c$ change with threshold $S$, we will give a detailed analysis based on the testing results of the two-stage assumption in subsection 4.4.

*4.3. Underlying mechanism of volatility evolution after extreme events*

The empirical results in Section 4.1 suggest that evolution behavior of volatility after extreme events differs in different stages, and thus the underlying mechanism in each stage may so differ. We explore the underlying mechanism by testing the proposed two-stage assumption based on the SAIH and MDH hypotheses in Section 2.3, i.e., investors sequentially transform from the uninformed state to the informed state in the first stage and informed investors subsequently dominate the market in the second stage. As shown in Fig. 3, the evolution behaviors of the volatility after endogenous and exogenous events are significantly different, hence we will present the testing results of the two-stage assumption for these two types of events respectively in the following. We also perform the ADF tests for the variables in Eqs. (6), (7) and (10) before the Granger-causality tests and volatility-volume relation tests, and the results show that all time series are stationary. To save space, we do not present the results of the ADF tests here.

*4.3.1. Test of two-stage assumption for volatility evolution after endogenous events*

We will first test the two-stage assumption for volatility evolution after endogenous events. The two stages of volatility evolution can be separated by the time constant $t_w$, which is the time it takes for all the traders to receive information. The values of $t_w$ for different thresholds $S$ obtained from the best fit of our VE model have been reported in Table 2 and we now test the assumption for the two stages $t \leq t_w$ and $t > t_w$ respectively.

In the first stage when $t \leq t_w$, we test the assumption that investors sequentially transform from the uninformed state to the informed state by examining the lead-lag relationships between the volatility and trading volume with Eqs. (6) and (7). We also examine whether informed investors dominate the market in this stage by testing the simultaneous relation between informed trading volume and the volatility with Eq. (10). The corresponding results are given in Tables 3 and 4.

The testing results in Table 3 reveal significant causal flows from trading volume to volatility for different thresholds $S$, suggesting that there exist lead-lag relations between trading volume and volatility. Hence, investors transform from the uninformed state to the informed state when $t \leq t_w$. In addition, the results in Table 4 reveal that $\beta_i$ for all different thresholds are insignificant

---

$t \leq t_w$ and a power law decay when $t > t_w$. This conclusion is valid for different $S$. We present the results of the optimal model with the smallest RSS in table 1.

**Table 3**
Granger-causality test between volatility and trading volume for the SSE 180 index after endogenous events when $t \leq t_w$.

|   | Null: volatility does not granger cause trading volume | | Null: trading volume does not granger cause volatility | |
|---|---|---|---|---|
|   | F-statistics | p-value | F-statistics | p-value |
| S = 2 | 2.41656 | 0.1511 | 22.4134 | 0.0018 |
| S = 4 | 8.15228 | 0.0115 | 10.1985 | 0.0057 |
| S = 6 | 1.64864 | 0.2400 | 31.4280 | 0.0008 |
| S = 8 | 2.36741 | 0.1625 | 22.4895 | 0.0015 |

This table reports F-statistics and p-values of F-test for the Granger-causality tests (Eqs. (6) and (7)) as $p=1$ and $q=1$, which are selected using the AIC criterion.

**Table 4**
Parameter estimates of the volatility-volume relation test for the SSE 180 index after endogenous events when $t \leq t_w$.

|   | $\beta_0$ | $\beta_i$ | $\beta_u$ | p-value |
|---|---|---|---|---|
| S = 2 | 0.0013 (0.0901) | -0.0015 (0.0730) | 0.0011 (0.0001) | 0.0003 |
| S = 4 | 0.0009 (0.4294) | -0.0010 (0.4174) | 0.0013 (0.0074) | 0.0151 |
| S = 6 | 0.0002 (0.9348) | -0.0003 (0.8988) | 0.0016 (0.0401) | 0.0540 |
| S = 8 | 0.0011 (0.9067) | -0.0016 (0.8511) | 0.0024 (0.1793) | 0.3495 |

The coefficient estimates reported in this table are from the estimation of Eq. (10). The p-values of t-test are given in parentheses beneath the coefficient estimates. The last column reports the p-values of F-test for the volatility-volume relation test using Eq. (10).

**Table 5**
Granger-causality between volatility and trading volume for the SSE 180 index after endogenous events when $t > t_w$.

|   | Null: volatility does not granger cause trading volume | | Null: trading volume does not granger cause volatility | |
|---|---|---|---|---|
|   | F-statistics | p-value | F-statistics | p-value |
| S = 2 | 2.4900 | 0.0648 | 1.5777 | 0.1997 |
| S = 4 | 2.4342 | 0.0694 | 1.2099 | 0.3103 |
| S = 6 | 2.0192 | 0.0829 | 1.1275 | 0.3513 |
| S = 8 | 1.5741 | 0.1018 | 0.7605 | 0.7222 |

This table reports F-statistics and p-values of F-test for the Granger-causality tests (Eqs. (6) and (7)) as $p=3$ and $q=3$, which are selected using the AIC criterion.

and $\beta_2$ for most of different thresholds are significant at the 5% significance level, indicating that informed investors do not dominate the market when $t \leq t_w$.

In the second stage when $t > t_w$, we test the assumption that informed investors dominate the market by using Eq. (10). To exclude the possibility that there exist investors transformed from uninformed state to informed state in this stage, we also examine the lead-lag relationships between the volatility and trading volume with Eqs. (6) and (7). The corresponding results are given in Tables 5 and 6.

The results in Table 5 show insignificant causality between trading volume and volatility at the 5% significance level for different thresholds $S$, suggesting that most investors transformed from the uninformed state to the informed state, which in turn implies that the process is likely to be completed at $t_w$. In Table 6, $\beta_i$ for different thresholds are all significant while $\beta_u$ for different thresholds are all insignificant at the 5% significance level, indicating that informed investors dominate the market when $t > t_w$.

Based on the testing results of two-stage assumption, we proceed our analysis to uncover the mechanism of volatility evolution after endogenous events as follows. In the first stage when $t \leq t_w$, uninformed investors who acquire no information totally dominate the market and make no trading decision at the moment when the endogenous event just happened. Hence the volatility relaxes toward equilibrium quickly. As time evolves, some investors acquire information from the change of price and transform from the uninformed state to the informed state. The informed investors and their decision-making behaviors increase over time, which directly leads to the increase of investors' trading behaviors. The market becomes more volatile, which gives rise to the increase of the resistance to recovery. As a consequence, the relaxation of volatility towards equilibrium becomes slower as $t$ increases in the first stage as shown in Fig. 3(a). In the second stage when $t > t_w$, the process in which investors transform from the uninformed state to the informed state is complete. The decision-making behavior therefore decreases over time, which may lead to the decrease of investors' trading behaviors. The market volatility becomes more restrained, which causes the decrease of the resistance to recovery. As a result, the volatility relaxes to equilibrium relatively faster in this stage than in the first stage as shown in Fig. 3(a).

### 4.3.2. Test of two-stage assumption for volatility evolution after exogenous events

We now test the two-stage assumption for volatility evolution after exogenous events. Since the exogenous events driven by news can be acquired quickly by investors, they can react and transform from the uninformed state to the informed state fast. As a result, $t_w$ almost becomes zero, which agrees with the fitted values of our VE model in Table 2. We now test the assumption that informed investors dominate the market when $t > 0$ by using equation (10). We also examine the lead-lag relationship between the volatility and trading volume with Eqs. (6) and (7) to exclude the possibility that there exists investors transformed from the uninformed state to the informed state in this stage. The corresponding results are given in Tables 7 and 8.

The testing results presented in Table 7 reveal insignificant causality between trading volume and volatility at 5% significance level for different thresholds $S$, suggesting that the process in which investors transform from the uninformed state to the informed state may be complete in a very short time after exogenous events, which gives $t_w \approx 0$. In Table 8, $\beta_i$ for different thresholds are all significant while $\beta_u$ for different thresholds are all insignificant at 5% significance level, suggesting that informed investors dominate the market when $t > 0$. All these results are similar to the case of endogenous events when $t > t_w$, as presented in Tables 5 and 6. The analysis is also similar to the case after endogenous events when $t > t_w$ in Section 4.3.1, and we will not go further into the details here.

### 4.4. Analysis of volatility evolution with VE model

In this subsection, we attempt to analyze volatility evolution after endogenous and exogenous events with our VE model. More specifically, a reasonable interpretation for the validity of our VE model and the tendency of estimated values of parameters $t_w$, $\rho$, $b$ and $c$ change with threshold $S$ will be offered based on the testing results of the two-stage assumption.

### 4.4.1. Analysis of volatility evolution after endogenous events with VE model

From the analysis of the assumption test, the resistance to recovery increases over time in the first stage when $t \leq t_w$, and the mean-reverting factor with a form $F_r \propto 1/t^c$ decreases slowly as time $t$ increases when $c > 1$. As a result, the volatility relaxes to equilibrium slowly as time $t$ increases, which agrees with the decay pattern of volatility in the first stage as shown in Fig. 3(a). In the second stage when $t > t_w$, the resistance to recovery decreases in this stage, and the mean-reverting factor with a form $F_r \propto 1/t$ decreases fast over time in comparison with the form in the first stage. Consequently, the volatility relaxes to equilibrium fast as

**Table 6**
Parameter estimates of the volatility-volume relation test for the SSE 180 index after endogenous events when $t > t_w$.

|       | $\beta_0$        | $\beta_i$       | $\beta_u$         | $p$-value               |
|-------|------------------|-----------------|-------------------|-------------------------|
| $S=2$ | -0.0010 (0.0001) | 0.0035 (0.0000) | -0.0018 (0.2846)  | $1.2237 \times 10^{-21}$ |
| $S=4$ | -0.0028 (0.0000) | 0.0048 (0.0000) | -0.0010 (0.1001)  | $2.3141 \times 10^{-16}$ |
| $S=6$ | -0.0026 (0.0001) | 0.0064 (0.0000) | -0.0027 (0.0711)  | $5.5394 \times 10^{-14}$ |
| $S=8$ | -0.0042 (0.0000) | 0.0059 (0.0000) | -0.0005 (0.5717)  | $5.5207 \times 10^{-7}$  |

The coefficient estimates reported in this table are from the estimation of Eq. (10). The $p$-values of $t$-test are given in parentheses beneath the coefficient estimates. The last column reports the $p$-values of $F$-test for the volatility-volume relation test using Eq. (10).

**Table 7**
Granger-causality between volatility and trading volume for the SSE 180 index after exogenous events.

|       | Null: volatility does not granger cause trading volume | | Null: trading volume does not granger cause volatility | |
|-------|--------------|----------|--------------|----------|
|       | F-statistics | p-value  | F-statistics | p-value  |
| $S=2$ | 1.8437       | 0.1258   | 2.0492       | 0.0926   |
| $S=4$ | 0.8381       | 0.5256   | 1.8894       | 0.1024   |
| $S=6$ | 1.9650       | 0.0900   | 1.3356       | 0.2551   |
| $S=8$ | 1.6180       | 0.1499   | 1.4961       | 0.1872   |

This table reports $F$-statistics and $p$-values of $F$-test for the Granger-causality tests (Eqs. (6) and (7)) as $p=3$ and $q=3$, which are selected using the AIC criterion.

time $t$ increases, in agreement with the steeper slopes of volatility curves when compared to the first stage as shown in Fig. 3(a).

In Table 2, the impact strength $\rho$ and parameter $b$ increases as the threshold $S$ increases in both stages. Since the behavior of the parameters $\rho$ and $b$ in the two different stages are consistent, we will analyze the results of both parameters in a unified picture. It is expected that extreme events with larger values of $S$, identified by the criteria $V(t) > S \langle V(t) \rangle_d$, generally have larger impact on stock markets. Hence, their impact strength $\rho$ is larger. Since events with greater impact attract more investors' attention, the investors transform faster from the uninformed state to the informed state. As a result, the faster the overall decision-making behavior increases, the faster the resistance to recovery increases, which then causes the faster decrease of the mean-reverting factor. A larger value of $b$ causes faster decrease of the mean reverting factor as time $t$ increases, which can explain the rational increase of estimated $b$ as the threshold $S$ increases.

One can also observe that the parameters $t_w$ and $c$ decrease as threshold $S$ increases and $c > 1$ in the first stage. It is well known that events with greater impact attract more investors' attention. Investors can therefore acquire information from the price change faster, and the investors also transform from the uninformed state to the informed state faster. Therefore $t_w$ becomes smaller as threshold $S$ increases. Thus, the larger the $S$ is, the faster the overall decision-making behaviors will increase and the faster the resistance to recovery increases. The mean-reverting factor will decrease faster, which can explain the decrease of the estimated values of parameter $c$ for larger threshold $S$.

*4.4.2. Analysis of volatility evolution after exogenous events with VE model*

In Table 2, $t_w = 0$ for different threshold $S$. Since the exogenous events driven by news can be acquired quickly, investors can react and transform from the uninformed state to the informed state fast. Hence, $t_w$ is close to zero, and the impact strength $\rho$ and parameter $b$ increases as threshold $S$ increases when $t > 0$. All these results are similar to the case of endogenous events when $t > t_w$. The analysis is also similar to the case after endogenous events when $t > t_w$ in Section 4.4.1, and we will not go further into the details here.

*4.5. Robustness and universality tests*

For the robustness test, we consider the volatility evolution of SSE 180 index during the period from December 16, 2013 to December 31, 2020. We also present the fitted results of our VE model for the volatility evolution after extreme events of the Hong Kong stock market to test the universality of our results.

*4.5.1. Robustness test*

To ensure that the previous findings are not artifacts due to the selection of our sampling period, we here consider a longer period from December 16, 2013 to December 31, 2020 for the robustness test. Figs. B.4 - B.6 and tables B.10 - B.11 in Appendix B.2 present the results of our VE model for volatility evolution of the SSE 180 index during this period.

One can see that the empirical results of volatility evolution after extreme events for the period from December 16, 2013 to December 31, 2020 are similar to the results for the sampling period from December 16, 2013 to November 22, 2016. In this period, the volatility evolution after extreme events also follows a stretched exponential decay in the initial stage and becomes a power law decay at later times, which can be well described by the VE model. Our findings for the volatility evolution of the SSE 180 index after extreme events are therefore robust.

**Table 8**
Parameter estimates of the volatility-volume relation test for the SSE 180 index after exogenous events.

|       | $\beta_0$        | $\beta_i$       | $\beta_u$        | $p$-value               |
|-------|------------------|-----------------|------------------|-------------------------|
| $S=2$ | -0.0023 (0.0000) | 0.0028 (0.0000) | 0.0003 (0.1732)  | $7.0027 \times 10^{-13}$ |
| $S=4$ | -0.0042 (0.0000) | 0.0045 (0.0000) | 0.0007 (0.0694)  | $1.0868 \times 10^{-14}$ |
| $S=6$ | -0.0033 (0.0000) | 0.0035 (0.0000) | 0.0011 (0.0770)  | $5.7472 \times 10^{-8}$  |
| $S=8$ | -0.0015 (0.0745) | 0.0022 (0.0031) | 0.0007 (0.0954)  | $5.0871 \times 10^{-4}$  |

The coefficient estimates reported in this table are from the estimation of Eq. (10). The $p$-values of $t$-test are given in parentheses beneath the coefficient estimates. The last column reports the $p$-values of $F$-test for the volatility-volume relation test using Eq. (10).

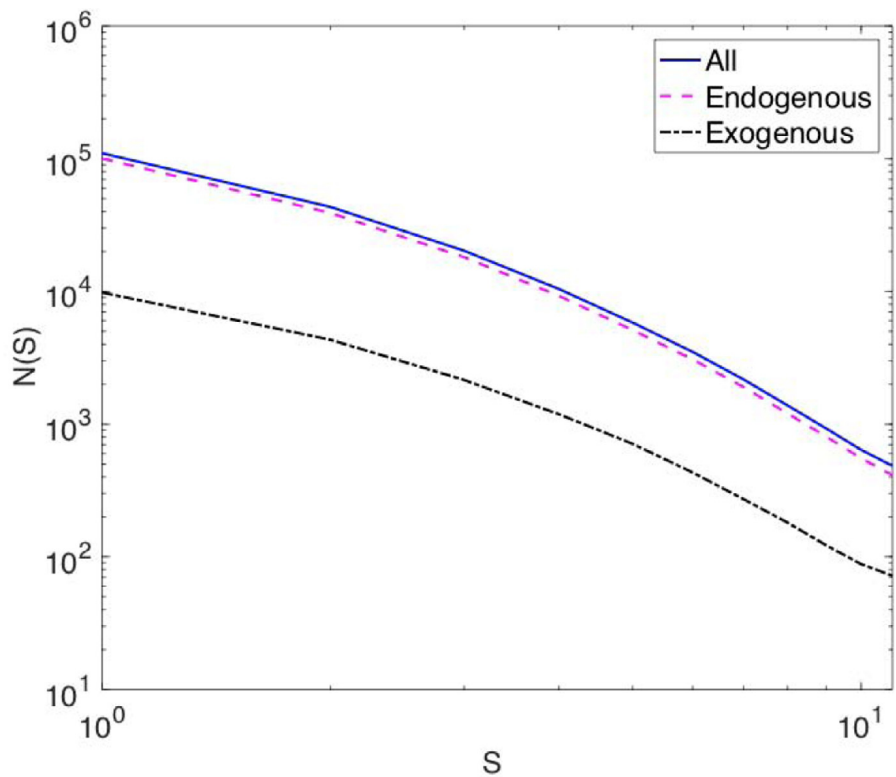

**Fig. B.4.** Numbers of all extreme events, endogenous events and exogenous events as a function of threshold $S$ for the SSE 180 index during the period from December 16, 2013 to December 31, 2020.

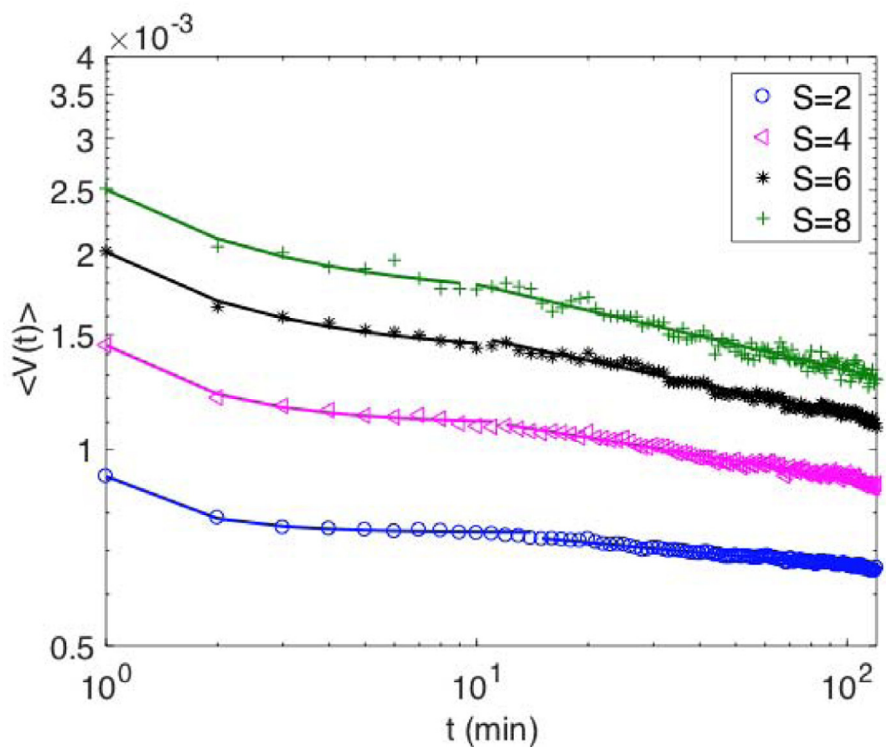

**Fig. B.5.** Volatility evolution after extreme events for threshold $S = 2, 4, 6, 8$ using the data of the SSE 180 index from December 16, 2013 to December 31, 2020. The solid lines in the figure are best fits to the data with Eq. (5).

**Table B.9**
Parameter estimates of VE model and Heston model for the volatility evolution of the SSE 180 index after extreme events for the period from December 16, 2013 to November 22, 2016.

| S | $t_w$ | $\rho e^{\frac{bt^{1-c}}{c-1}} (t \leq t_w)$ | | | $\rho t^{-b} (t > t_w)$ | | RSS |
|---|---|---|---|---|---|---|---|
| | | $\rho$ | b | c | $\rho$ | b | |

VE model ($\omega = 0$)

| S | $t_w$ | $\rho$ | b | c | $\rho$ | b | RSS |
|---|---|---|---|---|---|---|---|
| 2 | 13 | $0.89 \times 10^{-3}$ | 0.41 | 2.81 | $1.06 \times 10^{-3}$ | 0.05 | $3.17 \times 10^{-10}$ |
| 4 | 12 | $1.26 \times 10^{-3}$ | 0.42 | 2.37 | $1.73 \times 10^{-3}$ | 0.10 | $1.45 \times 10^{-9}$ |
| 6 | 10 | $1.65 \times 10^{-3}$ | 0.43 | 2.08 | $2.31 \times 10^{-3}$ | 0.12 | $2.49 \times 10^{-9}$ |
| 8 | 6 | $2.03 \times 10^{-3}$ | 0.46 | 1.88 | $2.75 \times 10^{-3}$ | 0.13 | $5.67 \times 10^{-9}$ |

VE model ($\omega > 0$)

| S | $t_w$ | $w + \rho e^{\frac{bt^{1-c}}{c-1}} (t \leq t_w)$ | | | $w + \rho t^{-b} (t > t_w)$ | | RSS |
|---|---|---|---|---|---|---|---|
| | | $\rho$ | b | c | $\rho$ | b | |
| 2 | 20 | $0.46 \times 10^{-3}$ | 0.45 | 2.52 | $0.62 \times 10^{-3}$ | 0.10 | $2.23 \times 10^{-9}$ |
| 4 | 20 | $0.83 \times 10^{-3}$ | 0.49 | 2.05 | $1.40 \times 10^{-3}$ | 0.16 | $2.12 \times 10^{-8}$ |
| 6 | 11 | $1.20 \times 10^{-3}$ | 0.49 | 2.03 | $2.00 \times 10^{-3}$ | 0.18 | $7.18 \times 10^{-8}$ |
| 8 | 6 | $1.70 \times 10^{-3}$ | 0.67 | 1.90 | $2.40 \times 10^{-3}$ | 0.18 | $1.88 \times 10^{-7}$ |

Heston model

| S | $\omega + e^{a-\theta t}$ | | RSS |
|---|---|---|---|
| | a | $\theta$ | |
| 2 | -8.2537 | 0.3506 | $9.62 \times 10^{-7}$ |
| 4 | -8.1551 | 0.5895 | $3.97 \times 10^{-6}$ |
| 6 | -8.1021 | 0.8226 | $9.13 \times 10^{-6}$ |
| 8 | -7.9400 | 1.0676 | $1.65 \times 10^{-5}$ |

The long-run equilibrium level w that is the average of volatility, equals to 0.0011.

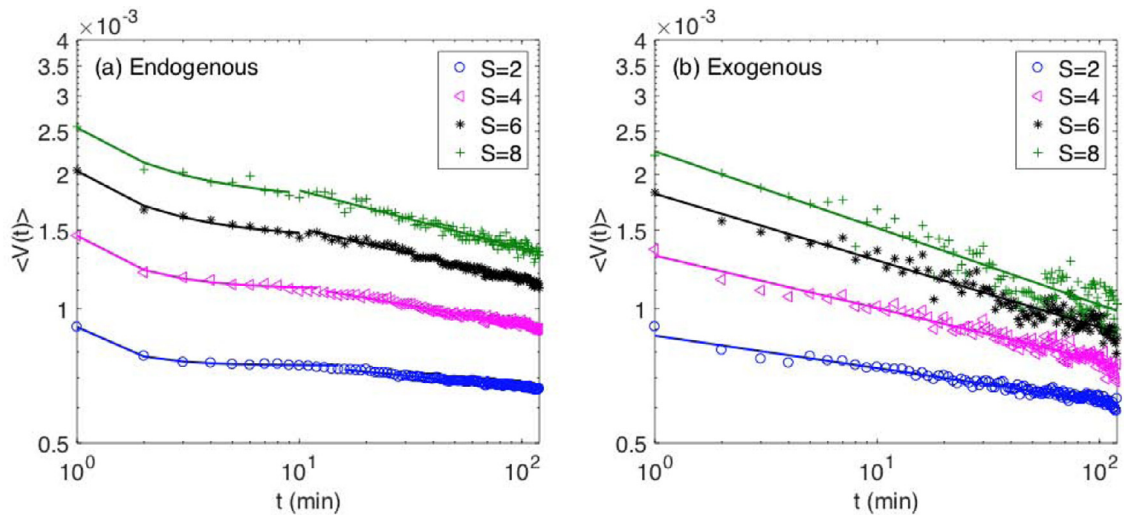

**Fig. B.6.** Volatility evolution after (a) endogenous and (b) exogenous events for threshold $S = 2, 4, 6, 8$ using the data of the SSE 180 index from December 16, 2013 to December 31, 2020. The solid lines in the figure are best fits to the data with Eq. (5).

**Table B.10**
Parameter estimates of VE model for the volatility evolution of the SSE 180 index after extreme events for the period from December 16, 2013 to December 31, 2020.

| S | $t_w$ | $\rho e^{\frac{bt^{1-c}}{c-1}} (t \leq t_w)$ | | | $\rho t^{-b} (t > t_w)$ | | RSS |
|---|---|---|---|---|---|---|---|
| | | $\rho$ | b | c | $\rho$ | b | |
| 2 | 14 | $0.74 \times 10^{-3}$ | 0.34 | 2.62 | $0.84 \times 10^{-3}$ | 0.05 | $1.43 \times 10^{-9}$ |
| 4 | 11 | $1.10 \times 10^{-3}$ | 0.35 | 2.19 | $1.40 \times 10^{-3}$ | 0.09 | $1.03 \times 10^{-8}$ |
| 6 | 10 | $1.40 \times 10^{-3}$ | 0.36 | 2.06 | $1.90 \times 10^{-3}$ | 0.12 | $2.97 \times 10^{-8}$ |
| 8 | 9 | $1.80 \times 10^{-3}$ | 0.37 | 2.05 | $2.40 \times 10^{-3}$ | 0.13 | $1.24 \times 10^{-7}$ |

This table reports the best-fit values of VE model as described by Eq. (5) for the volatility evolution of SSE 180 index after extreme events. The last column reports values of the residual sum of squares (RSS) for the best-fit of our VE model.

*4.5.2. Universality test*

To test the universality of empirical results presented above, we here study the volatility evolution in the Hong Kong stock market, one of the most developed markets in the world. The empirical results of the volatility evolution of HSI index after extreme events and the fitted results of our VE model are presented in Fig. B.7 and Table B.12 of Appendix B.

As shown in Fig. B.7 and Table B.12, the results of the volatility evolution after extreme events for the HSI index are consistent with those for the SSE 180 index of the Shanghai stock market. The volatility evolution after extreme events for the HSI index

**Table B.11**
Parameter estimates of VE model for the volatility evolution of the SSE 180 index after endogenous and exogenous events for the period from December 16, 2013 to December 31, 2020.

| | S | $t_w$ | $\rho e^{\frac{\lg^{1-c}}{c-1}}$ ($t \leq t_w$) | | | $\rho t^{-b}$ ($t > t_w$) | | RSS |
|---|---|---|---|---|---|---|---|---|
| | | | $\rho$ | $b$ | $c$ | $\rho$ | $b$ | |
| Endogenous | 2 | 15 | $0.74 \times 10^{-3}$ | 0.34 | 2.63 | $0.84 \times 10^{-3}$ | 0.05 | $1.51 \times 10^{-9}$ |
| | 4 | 12 | $1.10 \times 10^{-3}$ | 0.35 | 2.19 | $1.40 \times 10^{-3}$ | 0.09 | $1.04 \times 10^{-9}$ |
| | 6 | 10 | $1.40 \times 10^{-3}$ | 0.36 | 2.00 | $2.00 \times 10^{-3}$ | 0.12 | $3.03 \times 10^{-8}$ |
| | 8 | 9 | $1.70 \times 10^{-3}$ | 0.37 | 1.96 | $2.50 \times 10^{-3}$ | 0.13 | $1.40 \times 10^{-7}$ |
| Exogenous | 2 | 0 | | | | $0.87 \times 10^{-3}$ | 0.07 | $1.38 \times 10^{-8}$ |
| | 4 | 0 | | | | $1.30 \times 10^{-3}$ | 0.12 | $7.70 \times 10^{-8}$ |
| | 6 | 0 | | | | $1.80 \times 10^{-3}$ | 0.15 | $2.52 \times 10^{-7}$ |
| | 8 | 0 | | | | $2.30 \times 10^{-3}$ | 0.17 | $8.25 \times 10^{-7}$ |

This table reports the best-fit values of VE model as described by Eq. (5) for the volatility evolution of SSE 180 index after endogenous and exogenous events. The last column reports values of the residual sum of squares (RSS) for the best-fit of our VE model.

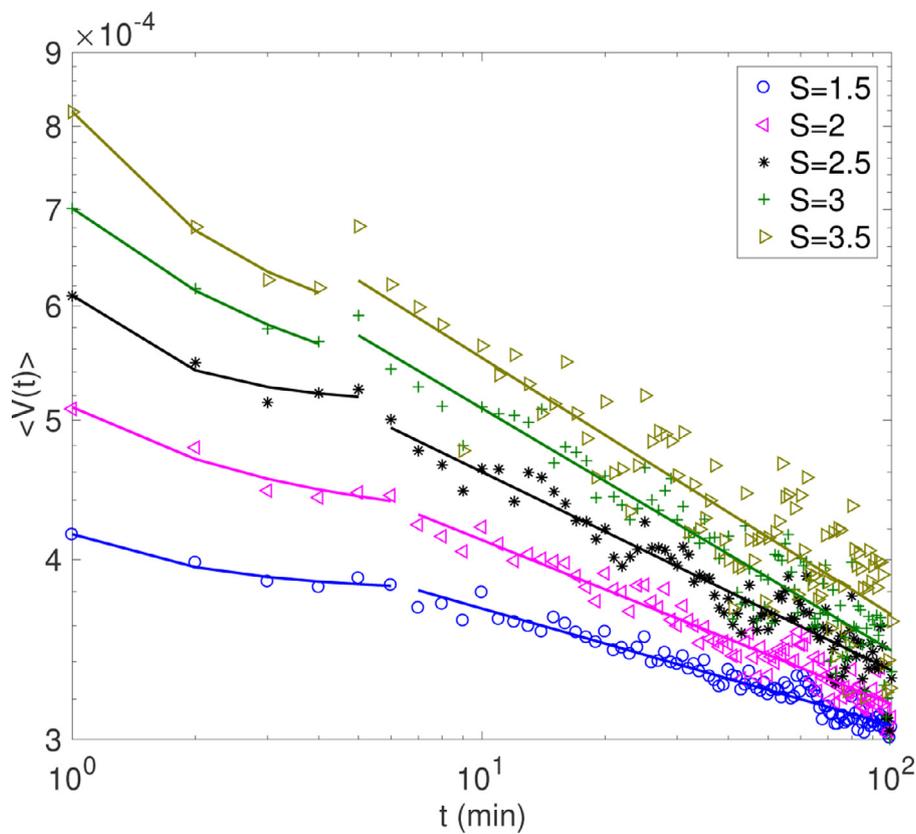

**Fig. B.7.** Volatility evolution after extreme events for threshold $S = 1.5, 2, 2.5, 3, 3.5$ using the data of the HSI index from January 11, 2016 to July 25, 2016. The solid lines in the figure are best fits to the data with Eq. (5).

**Table B.12**
Parameter estimates of VE model for the volatility evolution of the HSI index after extreme events for the period from January 11, 2016 to July 25, 2016.

| S | $t_w$ | $\rho e^{\frac{\lg^{1-c}}{c-1}}$ ($t \leq t_w$) | | | $\rho t^{-b}$ ($t > t_w$) | | RSS |
|---|---|---|---|---|---|---|---|
| | | $\rho$ | $b$ | $c$ | $\rho$ | $b$ | |
| 1.5 | 6 | $3.36 \times 10^{-4}$ | 0.07 | 2.21 | $4.50 \times 10^{-4}$ | 0.08 | $2.06 \times 10^{-9}$ |
| 2 | 6 | $3.79 \times 10^{-4}$ | 0.13 | 1.78 | $5.39 \times 10^{-4}$ | 0.12 | $5.93 \times 10^{-9}$ |
| 2.5 | 5 | $4.33 \times 10^{-4}$ | 0.17 | 1.50 | $6.32 \times 10^{-4}$ | 0.14 | $1.50 \times 10^{-8}$ |
| 3 | 4 | $5.29 \times 10^{-4}$ | 0.26 | 1.22 | $7.48 \times 10^{-4}$ | 0.16 | $3.13 \times 10^{-8}$ |
| 3.5 | 4 | $6.32 \times 10^{-4}$ | 0.57 | 1.04 | $8.02 \times 10^{-4}$ | 0.17 | $8.40 \times 10^{-8}$ |

This table reports the best-fit values of VE model as described by Eq. (5) for the volatility evolution of HSI index after extreme events. The last column reports values of the residual sum of squares (RSS) for the best-fit of our VE model.

shows similar behavior as compared to the SSE 180 index of the Shanghai stock market, which follows a stretched exponential decay in the initial stage and becomes a power law decay at later times, and can be well described by the VE model. The results obtained here imply that the evolutionary behavior of volatility after extreme events is universal.

## 5. Conclusion

In this paper, we propose a new dynamical model, the VE model based on the SV model to describe the two-stage volatility evolution of the SSE 180 index after extreme events. Differing from the SV model, an instantaneous impact and a revised mean-reverting factor are introduced in our VE model, in which the formula and relevant parameters are set according to investors' different behaviors in two stages. The fitted results of our new model show that the volatility evolution follows a stretched exponential decay in the initial stage and becomes a power law decay at later times, in good agreement with the empirical results. To further demonstrate the descriptive power of our VE model, we also use this model to fit the volatility evolution after endogenous and exogenous events separately. We find that the volatility evolution after endogenous events follows a stretched exponential decay in the first stage and a power law decay in the second stage, while the volatility evolution after exogenous events follows power law decays instantly after the impact. These results show a good match between the empirical results and the fitting results of our model for the volatility evolution after the extreme events including endogenous and exogenous events. One can also study the volatility evolution of the decomposed components of the index after extreme events by using the decomposition method introduced in [60].

We further explore the underlying mechanism that drives the volatility evolution of the SSE 180 index after extreme events by testing the proposed two-stage assumption based on the SAIH and MDH hypotheses. The testing results of the two-stage assumption indicate that investors transform from the uninformed state to the informed state in the initial stage and informed investors subsequently dominate the market after endogenous events, and investors dominate the market all the time after exogenous events. These findings are instructive for empirical studies on volatility evolution after extreme events, since they uncover the underlying mechanism of volatility evolution for different types of events. Based on these testing results, we offer an interpretation for the validity of the VE model and the estimated values of relevant parameters. We believe that our study gives a unified framework for understanding and explaining the volatility evolution after extreme events from the microscopic level of investor behaviors.

Our present study has applicable values for investors and policymakers. Since the dynamical model and the mechanism analysis in this paper can successfully describe and explain the volatility evolution after extreme events, it can therefore be applied to predict the stock price and volatility, and to describe the distribution of stock price returns, in particular after major shocks. By further identifying endogenous and exogenous events from the extreme events, the volatility could be predicted more precisely. In this respect, our study is useful to investors for risk management and portfolio selection, in particular for making timely and effective trading decisions in the initial stage after an extreme event. In the meantime, our research can help policymakers to understand the trend of the stock market after major shocks, and impose appropriate policies to avoid drastic fluctuations in the market. Our results of the underlying mechanism that investors transform from the uninformed state to the informed state and subsequently informed investors dominate the market will therefore be helpful to them when handling these financial issues. We also note that in an in-depth analysis of the relationship between volume and volatility, it is necessary to consider the intermediary role of liquidity [61], and we leave this for future research.

## Declaration of Competing Interest

None.

## CRediT authorship contribution statement

**Mei-Ling Cai:** Writing – review & editing, Data curation. **Zhang-HangJian Chen:** Writing – review & editing, Data curation, Software. **Sai-Ping Li:** Conceptualization, Methodology, Project administration. **Xiong Xiong:** Supervision, Funding acquisition. **Wei Zhang:** Supervision, Funding acquisition. **Ming-Yuan Yang:** Writing – original draft, Investigation, Software. **Fei Ren:** Supervision, Conceptualization, Methodology, Funding acquisition.

## Acknowledgements

This work was partially supported by the National Natural Science Foundation (Nos. 71790594 and 71871094), the Humanities and Social Sciences Fund sponsored by Ministry of Education of the People's Republic of China (No. 17YJAZH067), and the Fundamental Research Funds for the Central Universities (No. JKN02212304).

## Appendix A. Solution of the VE model

*A1. Long-run equilibrium value $\omega = 0$*

In the following, we provide details for obtaining the average solution of Eq. (4) in Section 2.2.

First, in Eq. (4), $F_e(t) = \rho \delta(t)$ denotes the instantaneous impact, $\eta(t)dt$ is the Gaussian white noise, and $\langle \eta(t) \rangle = 0$. Next, equation of the motion of the average volatility $\langle V(t) \rangle$ after an external instantaneous impact $F_e(t)$ is

$$\frac{d\langle V(t) \rangle}{dt} + \frac{b}{t^c}\langle V(t) \rangle = F_e. \tag{A.1}$$

We introduce the linear response theory [62], which can describe the decay of a system from a nonequilibrium state to an equilibrium state. If we assume the instantaneous impact $F_e(t) = \rho \delta(t)$ at time $t = 0$, then for time $t > 0$, the response is

$$\langle V(t) \rangle = \rho K(t), \tag{A.2}$$

where $K(t)$ is the linear response function.

Given that the average volatility $\langle V(t) \rangle$ obeys Eq. (A.1), we obtain the following equation for the response function

$$\rho \frac{dK(t)}{dt} + \rho \frac{b}{t^c} K(t) = F_e. \tag{A.3}$$

Since the instantaneous impact $F_e(t) = \rho \delta(t)$ is infinite at $t = 0$ and equals zero at $t > 0$, thus we have

$$\frac{dK(t)}{K(t)} = -\frac{b}{t^c} dt, \tag{A.4}$$

when time $t > 0$.

Integrating both sides of Eq. (A4), we get

$$\ln K(t) = \begin{cases} \frac{bt^{1-c}}{c-1} & c > 1 \\ -b \ln t & c = 1 \end{cases}, \tag{A.5}$$

and the response function $K(t)$ is given by

$$K(t) = \begin{cases} e^{\frac{bt^{1-c}}{c-1}} & c > 1 \\ t^{-b} & c = 1 \end{cases}. \tag{A.6}$$

Finally, from Eq. (A2), the average solution of Eq. (4) is

$$\langle V(t) \rangle = \rho K(t) = \begin{cases} \rho e^{\frac{bt^{1-c}}{c-1}} & t \leq t_w \\ \rho t^{-b} & t > t_w \end{cases}. \tag{A.7}$$

*A2. Long-run equilibrium value $\omega > 0$*

If the volatility is defined as the standard deviation of stock return, the volatility would decay to its long-run equilibrium level and $\omega$ would be greater than 0. We here provide the average solution of the VE model when the long-run equilibrium value $\omega > 0$. Corresponding to the SV model, $F_r(t)$ would take the form

$$F_r(t) = \frac{b}{t^c}(w - V(t)), \tag{A.8}$$

and the volatility $V(t)$ can be expressed as

$$dV(t) = \rho \sigma(t) dt + \frac{b}{t^c}(w - V(t)) + \eta(t), \tag{A.9}$$

Eq. (A10) can be solved exactly and the average solution is

$$\langle V(t) \rangle = \begin{cases} \omega + \rho e^{\frac{bt^{1-c}}{c-1}} & t \leq t_w \\ \omega + \rho t^{-b} & t > t_w \end{cases}. \tag{A.10}$$

The form in Eq. (A9) ensures that $\frac{dV(t)}{dt}$ goes to 0 and the volatility $V(t)$ converges to its long-run equilibrium level $\omega$ when $t \to \infty$. We take the long-run equilibrium level $\omega$ as the average of volatility during the sampling period, which equals to 0.0011. Table B.9 presents the results of the VE model with a long-run equilibrium level $\omega > 0$ for volatility evolution of the SSE 180 index after extreme events.

**Appendix B. Numerical results**

*B1. Results of the VE model and the Heston model*

*B2. Results of robustness and universality tests*